\documentclass[twocolumn]{aastex631}
\usepackage{lineno}
\usepackage[utf8]{inputenc}
\usepackage{graphicx}
\usepackage{booktabs}
\usepackage{threeparttable, tablefootnote}
\usepackage{float}

\begin{document}
\title{An Increase in the Prevalence of Ionized Outflows in Galaxies with Coronal Line Emission: Feedback in Action?}
\author[0000-0012-3245-1234]{William Matzko}
\affiliation{George Mason University, Department of Physics and Astronomy, MS3F3, 4400 University Drive, Fairfax, VA 22030, USA}

\author[0000-0003-2277-2354]{Shobita Satyapal}
\affiliation{George Mason University, Department of Physics and Astronomy, MS3F3, 4400 University Drive, Fairfax, VA 22030, USA}

\author[0000-0003-4701-8497]{Michael Reefe}
\altaffiliation{National Science Foundation, Graduate Research Fellow}
\affiliation{MIT Kavli Institute for Astrophysics and Space Research, Massachusetts Institute of Technology, Cambridge, MA 02139, USA}

\author[0000-0002-0913-3729]{Jeffrey McKaig}
\affiliation{George Mason University, Department of Physics and Astronomy, MS3F3, 4400 University Drive, Fairfax, VA 22030, USA}

\author[0000-0003-3432-2094]{Remington O. Sexton}
\affiliation{George Mason University, Department of Physics and Astronomy, MS3F3, 4400 University Drive, Fairfax, VA 22030, USA}
\affiliation{U.S. Naval Observatory, 3450 Massachusetts Avenue NW, Washington, DC 20392-5420, USA}

\author[0000-0003-3152-4328]{Sara Doan}
\affiliation{George Mason University, Department of Physics and Astronomy, MS3F3, 4400 University Drive, Fairfax, VA 22030, USA}

\begin{abstract}

Coronal lines (CLs), which arise from collisionally excited forbidden transitions from highly ionized species, are a powerful diagnostic tool in uncovering active galactic nuclei (AGNs) and constraining their properties. However, recent optical surveys are finding that coronal lines are rarely detected in the majority of local AGNs, possibly as a result of the depletion  of elements from the interstellar gas onto dust grains. Prominent CL emission may therefore only arise when dust is being destroyed in the highly ionized gas in AGNs. To explore the possibility that dust destruction may be caused by ionized gas outflows in galaxies with prominent CLs, we present the first large-scale systematic study of ionized outflows, as traced by the [O III] $\lambda$5007 emission, in galaxies displaying CL emission relative to a robust control sample of non-CL-emitting galaxies. We find: 1) galaxies that display CL emission have a significantly elevated outflow incidence rate compared to their matched controls, 2) the outflow luminosity is significantly higher in the CL-emitters, 3) the CL-emitters have systematically lower intrinsic extinction toward the ionized gas compared with the controls, 4) there are significant correlations between the CL luminosity and outflow velocity for the iron CLs, with similar relationships found between the CL FWHM and outflow luminosity. These observations are consistent with dust destruction in an outflowing wind from a dusty torus causing efficient CL emission.

\end{abstract}

\section{Introduction} \label{sec: intro}
Identifying accreting supermassive black holes (SMBHs) in galaxies and characterizing their properties is essential in order to constrain theories of SMBH origins, their growth, and their co-evolution with galaxies. However, uncovering active galactic nuclei (AGNs) across the full range of galaxy properties in local and high-redshift galaxies poses a significant challenge, as accretion signatures are frequently missed in optical spectroscopic, broad-band mid-infrared, X-ray, and radio surveys due to obscuration of the nuclear region or contamination by emission from star formation activity in the host galaxy \citep[e.g.,][]{condon1991,2009MNRAS.398.1165G,2015ApJ...811...26T, 2018ApJ...858...38S}.
\par

The advent of JWST has underscored how identifying AGNs and accurately constraining their properties has become even more challenging than before. The overabundance of low luminosity broad-line selected AGNs discovered by JWST at redshifts from z~$\approx$ 4 - 11 \citep[e.g.,][]{2024ApJ...964...39G, 2023arXiv230311946H,2023ApJ...954L...4K,2023ApJ...953L..29L,  2024Natur.627...59M, 2024A&A...691A.145M, 2024ApJ...963..129M, 2024Natur.628...57F, 2024MNRAS.531..355U}, potentially with black hole masses comparable to their host galaxies \citep[e.g.,][]{2024A&A...691A.145M, 2024arXiv240303872J, 2023arXiv230311946H}, would have significant consequences to theories of SMBH origins and their co-evolution with galaxies \citep[e.g.,][]{2024ApJ...960L...1N, 2024ApJ...964..154P}. However, despite their optical broad line signatures, the vast majority do not show the hallmark signatures of an AGN: they are undetected in the X-rays \citep{2024ApJ...969L..18A,2024arXiv240303872J,2024arXiv240500504M, 2024arXiv240413290Y}, do not show variability \citep{2024arXiv240704777K,2024arXiv241205242T}, the subset known as ``little red dots'' (LRDs) \citep{2024ApJ...963..129M} are unusually compact and have red UV to optical colors suggestive of dust obscuration \citep[e.g.,][]{2024ApJ...964...39G, 2024ApJ...968...38K} but they have a puzzling blue excess and lack the steep IR continuum characteristic of hot torus emission \citep[e.g.,][]{2024ApJ...968....4P,2024arXiv240302304W, 2024ApJ...968...34W}, leading some to suggest that not all members of this newly discovered population are indeed AGNs \citep[e.g.,][]{2024ApJ...977L..13B}.
\par
High-ionization forbidden lines, commonly called "coronal lines" (CLs) due to their initial discovery in the solar corona, are not typically produced even by the most vigorous star formation activity and are therefore an unambiguous diagnostic for identifying AGNs and analyzing their properties \cite[e.g.,][]{2021ApJ...906...35S}. However, although the AGN accretion disk should produce a copious supply of photons capable of producing these high-ionization lines, many of the high-z broad line objects discovered by JWST do not show any CLs in their rest frame optical/UV spectra and only a few show marginal detections, indicating the line luminosities are low \citep[e.g.,][]{2023ApJ...954L...4K, 2023A&A...677A.145U, 2024arXiv240913047L}. In addition, recent systematic surveys of the optical CLs in large samples of local galaxies are revealing that optical CL emission is extremely rare in the general galaxy population given current survey sensitivities, even in galaxies classified as AGNs based on standard narrow line ratio diagnostics \citep{2021ApJ...922..155M, 2021ApJ...920...62N, Reefe2022,2023ApJ...945..127N}. These studies also show that CL emission is preferentially found in galaxies with the least dust extinction \citep{2023ApJ...945..127N}, suggesting that the presence of dust may play a role in suppressing the emission. 
\par

The presence of dust in the high-density ionized gas in AGNs is expected to impact the emission line spectrum in a number of ways. First, the presence of dust modifies the chemical composition of the gas phase due to the depletion of refractory elements onto the grains. In addition, grains impact the heating and cooling of the gas, and can absorb ionizing radiation that would have otherwise been absorbed by the gas.
 Recent photoionization models by \citet{McKaig2024} demonstrate that gas phase depletion of refractory elements onto dust grains and absorption of ionizing photons by dust  has a dramatic effect on the CL emission, suppressing the line luminosity by as much as 3 orders of magnitude compared with dust-free gas. Prominent CL emission may therefore arise only when there is sufficient grain destruction in the highly-ionized gas, a result that has been noted in several other works as well \citep[e.g.,][]{1997ApJS..110..287F,2003AJ....125.1729N,2006A&A...447..863N} . Indeed, recent studies have shown that the most commonly detected CLs are from non-refractory elements such as Neon \citep{2021ApJ...920...62N, Reefe2022,Reefe2023,2023ApJ...945..127N}, as one would expect if the highly refractory elements are typically  depleted from the gas phase onto dust grains. Dust may play a role at suppressing CL emission even at high redshift. Recent JWST observations have revealed a significant dust content in both AGNs \citep[e.g.,][]{2023ApJ...956...61A, 2024arXiv241007340B, 2024arXiv240710760L, 2024ApJ...968...38K, 2024MNRAS.534.2585V} and star forming galaxies \citep[e.g.,][]{2024arXiv241023959B, 2024A&ARv..32....2S, 2024ApJ...968...34W, 2023Natur.621..267W}. 
\par

If the cause of the weak high-ionization emission lines in local and high redshift galaxies is due in part to depletion of elements from the gas phase onto dust grains, the destruction of the dust in the high density gas around AGNs must be taking place in AGNs displaying prominent CLs. Outflows in the ionized gas are a possible mechanism for dust destruction in shocked gas through kinetic sputtering or grain-grain collisions. Such a finding would be consistent with the finding that the CL detection fraction appears to be higher in quasars that display outflows in the [\ion{O}{3}]~$\lambda$5007 line \citep{2025arXiv250117067D}, and that the CLs themselves are often blueshifted and likely associated with energetic outflows \citep[e.g.,][]{1997A&A...323..707E,2006ApJ...653.1098R,2011ApJ...743..100R}. An association between CLs and outflows has previously been reported by \citet{2021ApJ...911...70B} in dwarf galaxies, suggesting that CL emission is enhanced in galaxies with outflows, possibly as a result of grain destruction in shocks.
These studies suggest that prominent CL emission may be associated with powerful ionized gas outflows in AGNs, and that galaxies displaying prominent CL emission may capture a population of galaxies undergoing enhanced AGN feedback. While these results are suggestive of an association of prominent CL emission and outflows, the outflow incidence and CL detections are both affected by selection biases since the detection of both a CL line and an outflow depends on a number of factors including the sensitivity of the observations, and the presence, classification, and luminosity of an AGN. No systematic study of the relationship between outflows and CL emission has thus far been carried out.
\par
In this work, we carry out the first systematic study of the incidence and properties of ionized outflows, as traced by the [OIII]~$\lambda$5007 emission line, in a large sample of CL emitters identified in the Sloan Digital Sky Survey (SDSS) relative to a carefully matched control sample. This paper builds on our previous work on the  The Coronal Line Activity Spectroscopic Survey (CLASS) in which the detection fractions and properties of the twenty optical coronal lines in a sample of almost 1 million galaxies is explored \citep{Reefe2022,Reefe2023}. The goal of this work is to determine if the incidence of outflows is enhanced in CL emitters compared to matched controls, and if there are any statistically significant trends between outflow properties, dust extinction, and CL emission. 
\par
This paper is organized as follows. In Section \ref{sec:sample}, we describe the CLASS sample and our criteria for the selection of our control sample. In Section \ref{sec:Spectral Fitting and Outflow Criteria}, we describe our methodology for spectral fitting and outflow criteria. In Section \ref{sec: results}, we discuss the incidence of ionized outflows relative to our controls and any statistical trends in outflow properties, CL emission, and dust extinction. We discuss possible interpretations for our results in Section \ref{sec: discussion} and summarize our main findings in Section \ref{sec: conclusion}. Throughout this work, we adopt a flat $\Lambda$CDM cosmological model with $H_0 = 70$ km~s$^{-1}$~Mpc$^{-1}$, $\Omega_m = 0.3$, and $\Omega_\Lambda = 0.7$.

\section{Sample Selection} \label{sec:sample}
\subsection{Coronal Line Sample} \label{subsec: CL_sample}

For this work, we use the CLASS catalog of CL-emitters described in our previous work \citep{Reefe2022, Reefe2023}. In this way, we ensure that the CL properties of all galaxies in the sample are analyzed uniformly with the same methodology employed for spectral fitting and the same criteria applied for all reported CL detections. We refer the reader to our previous work for a detailed description of the spectral fitting and analysis used in the construction of the catalog. In brief, the CLASS galaxy sample is drawn from the  SDSS MPA/JHU DR8 catalog\footnote{\href{https://wwwmpa.mpa-garching.mpg.de/SDSS/DR7/}{https://wwwmpa.mpa-garching.mpg.de/SDSS/DR7/}}, which contains 952,138 optical ($\sim$3300--8000 \AA) spectra of galaxies. The sample of CL-emitters is based on a systematic search for every CL that is present at optical wavelengths, from the [\ion{Ne}{5}] $\lambda\lambda$3346,3426 doublet at the blue end to the high ionization-potential [\ion{Fe}{11}] $\lambda$7892 at the red end (a total of twenty CLs). The resulting sample consists of 258 galaxies. We discard one of the 258 CL-emitters because its high redshift ($z \sim 0.9$) places the [OIII]~$\lambda$5007 emission line outside the SDSS wavelength range, resulting in a final sample size of 257 galaxies. The galaxies from this sample are local ($z \lesssim 0.3$), have a stellar mass less than $\sim 10^{11}~\mathrm{M_{*}/M_{\odot}}$, have an elevated SFR $\sim1\sigma$ above the $M_*-SFR$ main sequence \citep{McGaugh2017}, and have $\mathrm{[O~III]}~\lambda5007$ luminosities greater than $\gtrsim 10^{41}~\mathrm{erg}~\mathrm{s}^{-1}$.
\par
Our goal in this work is to compare outflow properties in CL emitters to those in galaxies without CLs, independent of any selection biases. Because outflow properties are known to depend on the the presence of an AGN and its classification based on optical narrow line diagnostics or mid-infrared color selection \citep[e.g.,][]{Woo2016, Lutz2020, Avery2021, Matzko2022}, we construct 4 sub-samples of CL galaxies based on these classifications in which the outflow incidence and properties are explored and compared to non-CL galaxies in each sub-class. Each of the 4 subclasses are constructed as follows. Galaxies in our CL sample are classified based on optical narrow line ratios using the standard Baldwin-Phillip-Terlevich (BPT) diagnostics \citep{Baldwin1981} and mid-infrared color selection. We use the standard K01 \citep{Kewley2001} and K03 \citep{Kauffmann2003} optical selection criteria to classify galaxies as K01 AGN, K03 AGN, and K03 SF (hereafter SF) using fluxes from the full CLASS catalog \citep{Reefe2023}. For mid-infrared selection, we utilize the Wide-field Infrared Survey Explorer (WISE, \citet{Wright2010}) magnitudes, W1 and W2 at 3.4 and 4.6 microns respectively, from the full CLASS catalog (originally obtained from the ALLSKY WISE source catalog\footnote{https://wise2.ipac.caltech.edu/docs/release/allwise/}). The WISE sample is composed of objects that satisfy the WISE color cut $W1 - W2 > 0.5$, which has been shown to provide a more complete sample of AGNs at $z \lesssim 0.5$ \citep{Blecha2018}. Finally, for the SF sub-sample, we also require galaxies to have a WISE magnitude of $W1-W2 < 0.5$ to remove potential AGN contamination, since optical narrow line diagnostics may not detect some AGNs, especially in galaxies with high dust content \citep{Satyapal2008, Satyapal2014, Truebenbach2017}.


\begin{deluxetable*}{ccccccc}[ht]

\tablecaption{General properties of each sub-sample. Quoted quantities are the mean values for each sub-sample. Outflow uncertainties are calculated with binomial statistics, with upper/lower uncertainties corresponding to a 90\% confidence level.  All other uncertainties are calculated using a 1$\sigma$ standard deviation.}

\label{table:general}
\tablenum{1}

\tablehead{\colhead{Name} & \colhead{Total Number} & \colhead{z} & \colhead{ Outflow Fraction}  & \colhead{[O III] Lum} & \colhead{Stellar Mass} & \colhead{SFR} \\ 
\colhead{} & \colhead{} & \colhead{} & \colhead{(\%)} & \colhead{(Log$_{10}$(erg s$^{-1}$))}  & \colhead{(Log$_{10}$(M$_{\odot}$))} & \colhead{(Log$_{10}$(M$_{\odot}$ yr$^{-1}$))} } 

\startdata
CLASS SF & 24 & 0.102 $\pm$ 0.098& 16.67$^{13.92}_{9.19}$ & 41.20 $\pm$ 0.83 & 10.70 $\pm$ 0.77 & 0.899 $\pm$ 1.453 \\
Control SF & 72 & 0.093 $\pm$ 0.060 & 8.33$^{5.84}_{3.90}$ & 41.96 $\pm$ 1.24 & 10.61 $\pm$ 0.35 & 0.568 $\pm$ 0.745 \\
\rule{0pt}{3ex} CLASS K03 & 83 & 0.0979 $\pm$ 0.0619 & 78.31$^{5.82}_{6.90}$ & 41.62 $\pm$ 0.53 & 10.76 $\pm$ 0.41 & 0.333 $\pm$ 0.419 \\
Control K03 & 249 & 0.0886 $\pm$ 0.0557 & 63.85$^{4.01}_{4.18}$ & 41.55 $\pm$ 0.89 & 10.61 $\pm$ 0.35 & 0.692 $\pm$ 0.841 \\
\rule{0pt}{3ex} CLASS K01 & 80 & 0.0958 $\pm$ 0.0611 & 83.75$^{5.20}_{6.55}$ & 41.62 $\pm$ 0.50 & 10.86 $\pm$ 0.42 & 0.377 $\pm$ 0.463 \\
Control K01 & 240 & 0.0824 $\pm$ 0.0664 & 70.00$^{3.86}_{4.11}$ & 41.55 $\pm$ 1.04 & 10.67 $\pm$ 0.53 & 0.388 $\pm$ 0.662 \\
\rule{0pt}{3ex} CLASS WISE & 88 & 0.0983 $\pm$ 0.0618 & 73.86$^{6.10}_{6.95}$ & 41.81 $\pm$ 0.52 & 10.57 $\pm$ 0.42 & 0.535 $\pm$ 0.736 \\
Control WISE & 264 & 0.120 $\pm$ 0.049 & 60.98$^{3.96}_{4.09}$ & 42.19 $\pm$ 1.35 & 10.73 $\pm$ 0.28 & 0.572 $\pm$ 0.600
\enddata

\end{deluxetable*}


\subsection{Control Selection} \label{subsec: control_sample}
A control sample of galaxies without CLs was constructed using the full CLASS catalog for each of the CL galaxies in each of the 4 sub-samples in order for their outflow properties can be differentially assessed, independent of any selection biases. First, controls are simultaneously matched in redshift ($\pm 0.01$), stellar mass ($\pm 0.1$ dex), and [O III] $\lambda$5007 luminosity ($\pm0.5$ dex), all of which may show a dependence on outflow and coronal line properties \citep[e.g.][]{Matzko2022}. In addition, to account for variations in the signal-to-noise ratio (SNR) of the SDSS spectra in the pool of control galaxies without CL detections, we exclude observations in which the sensitivity of the spectrum is not sufficient to enable a 3$\sigma$ detection of the given CL in the CL sample. This ensures that our sample of control galaxies without CLs are genuine non-detections, given the sensitivity of the corresponding CL emitter to which it is matched. In other words, for each CL in the CL galaxies in our sample, the matched control galaxies should have had a CL detection for each observed CL, had it been present. We require all CL-emitters to be matched with exactly three \textit{unique} controls that are of the same optical/infrared classification (e.g. K01 CL emitters are matched only to K01 controls, etc.); CL-emitters that could not be matched with three unique controls were discarded from the analysis. While we ensure each \textit{sub-sample} contains unique CL-emitters and matched controls, the controls \textit{across} sub-samples may not be unique. Our matching criteria leaves us with a final sample of 24 SF, 80 K01 AGN, 83 K03 AGN, and 88 WISE AGN CL-emitters. The general properties of each of our sub-samples is summarized in Table \ref{table:general}.

\section{Spectral Fitting Procedures and Outflow Criteria}\label{sec:Spectral Fitting and Outflow Criteria}

We use the [O III] $\lambda$5007 emission line to trace outflows in all galaxies in our sample. This line is one of the strongest features in the optical spectrum of all emission line galaxies, and is located in a wavelength region free from strong stellar absorption features. Since it is produced in the lower density narrow line region, any asymmetry and broadening of the profile can often be attributed to large-scale winds. For these reasons, it is commonly used in the literature to search for and characterize ionized outflows \citep[e.g.][]{Wylezalek2020, Matzko2022}

While the full CLASS catalog includes fits for outflows in the [O III] $\lambda$5007 line, we re-fit the [O III]/H$\beta$ complex with a more careful line-testing algorithm to ensure secondary outflowing components are robustly detected. Spectral fitting is performed with a custom version of the open-source \textsc{PYTHON} code Bayesian AGN Decomposition Analysis for SDSS Spectra (\textsc{BADASS}) based on version 9.3.0. A full description of the code can  be found in \citet{Sexton2021}. Here, we provide a brief summary of the fitting algorithm and input parameters. \textsc{BADASS} works by modeling various spectral components (e.g. emission lines, absorption lines, AGN power-laws, etc.) and sequentially subtracting them from the data, leaving only the stellar continuum. The stellar continuum is fit with either empirical stellar templates or a single stellar population model, depending on the SNR of the spectrum. Emission lines may be fit with a variety of line profiles, but here we use a Gaussian profile that is fit for amplitude, dispersion ($\sim$FWHM/2.355), and velocity offset from the expected rest wavelength of the line. Initial fits to the spectra are obtained using the Basinhopping global optimization algorithm from the standard Python library \textsc{SCIPY.OPTIMIZE.BASINHOPPING}, with local minimization performed using the \textsc{SLSQP} algorithm. The best fit parameters from this process are then used as the initial guesses for a Markov Chain Monte Carlo (MCMC) algorithm, implemented by the \textsc{EMCEE} package \citep{ForemanMackey2013}, to estimate robust parameter uncertainties. 

In order to test for the presence of outflows, we fit the [O III] $\lambda$5007 line with and without a secondary Gaussian outflow component using the same process described above (excluding MCMC iterations). We utilize a bootstrap technique, injecting Gaussian noise into the spectrum at each iteration for 10 iterations, and comparing the resultant fit distributions using both an A/B test and an F-test. If the corresponding p-values for the A/B test \textit{or} F-test are greater than 0.9, we assume a secondary component is justified and include it in the final model produced by \textsc{EMCEE}. Finally, to ensure the final fits are robust, we require that the SNR of the outflowing component be at least 3, and the FWHM of the outflowing component  be greater than the FWHM of the narrow component within their uncertainties to a 3$\sigma$ level. Outflow fits that do not meet this criteria are assumed to be non-detections. 

\section{Results} \label{sec: results}

\subsection{Outflow Incidence} \label{subsec: outflow_incidence}


\begin{figure}[!t]
\includegraphics[width = 0.45\textwidth]{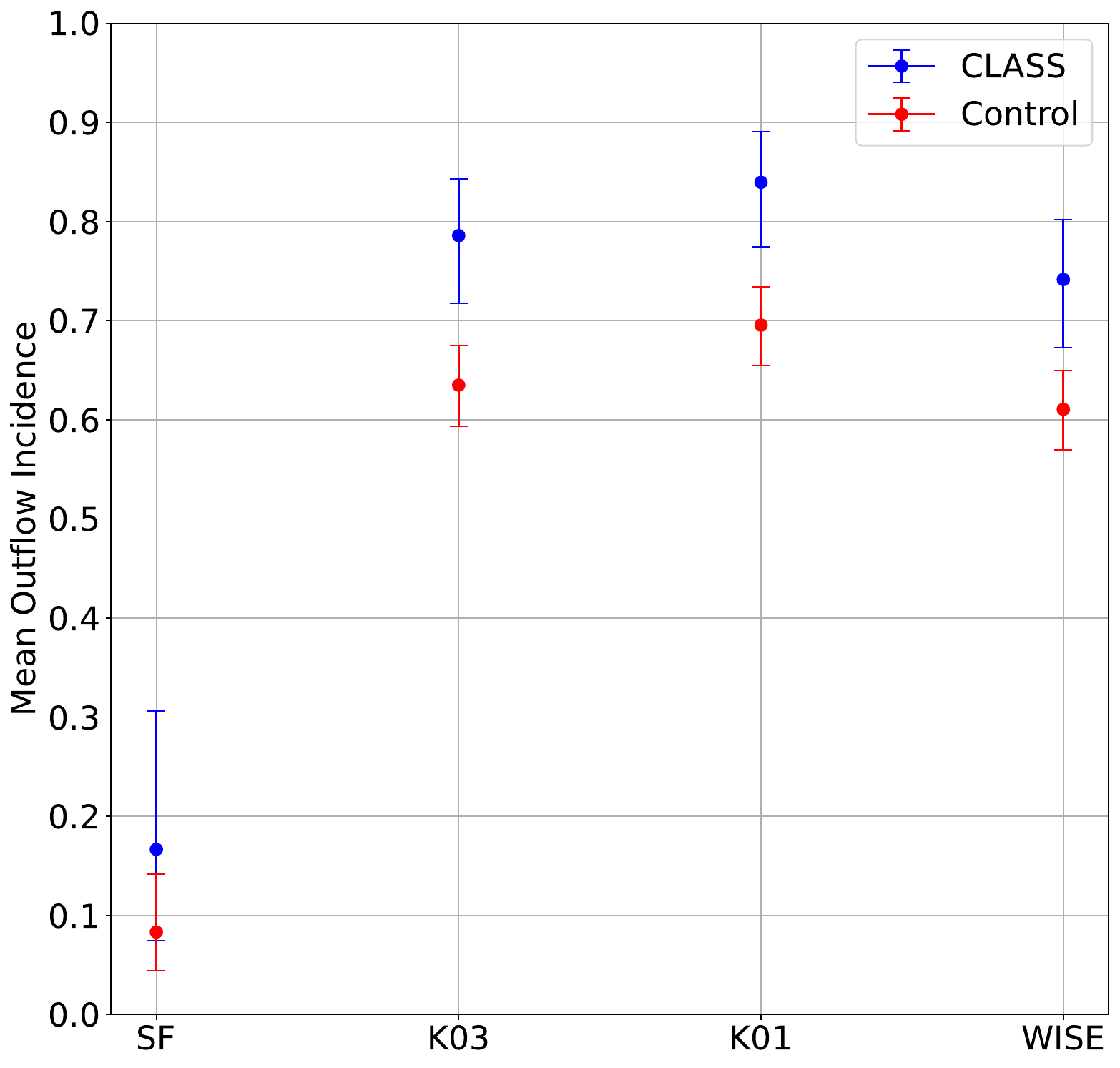}
\caption{Average outflow incidence in CL emitters (CLASS sample, blue) vs. matched control sample (red) for each sub-sample. Error bars are calculated using binomial counting statistics and correspond to a 90\% confidence level. In each AGN sub-sample, the outflow incidence in the CL-emitters is significantly higher than in the matched controls.}
\label{fig:incidence}
\end{figure}


Figure \ref{fig:incidence} shows the average [O III] $\lambda$5007 outflow incidence rate in each of our sub-samples. Here, the uncertainties are computed using binomial statistics with errors representing the 90\% confidence level. Figure \ref{fig:incidence} demonstrates that the outflow incidence is systematically higher in CL emitters verses their matched controls for all 4 sub-samples. For galaxies classified as AGNs through any of the 3 classifications, the difference in the outflow incidence between the CL-emitters and their matched controls is significant at the 90\% level, with the K01 and K03 AGN differences being significant at the 95\% level.

\par

Figure \ref{fig:incidence} also shows that the outflow incidence for galaxies classified as AGNs is high in our AGN sub-samples, roughly above 70 percent, while for SF galaxies the incidence is only about 10 percent. Outflow incidence rates vary widely across the literature, due to the vastly different sample constructions and different methods of detecting (and defining) outflows. Outflow incidences reported in various studies have ranged from $\sim$10-30 percent \citep[e.g.][]{Manzano2019, Wylezalek2020, Avery2021, Kukreti2023} to as high as $\sim$40-95 percent \citep[e.g.][]{Veilleux2013, Perna2017, Toba2017, Rakshit2018}. The range in outflow fractions reported in the literature is due to a number of a factors, including the stellar mass of the galaxy, the presence of an AGN, its classification, and the luminosity of the galaxy. In particular, the (bolometric) luminosity has been found to correlate strongly with outflow incidence \citep[e.g.][]{Woo2016, Matzko2022}, with an [O III] luminosity $\gtrsim$10$^{41}$ erg s$^{-1}$ being an approximate threshold for significant increases in outflow detection; at these higher luminosities, outflow detection fractions can be $\approx$ 60 - 70 percent or more. Further, \citet{Forster2019} found a correlation between the stellar mass of a galaxy and outflow incidence (see their Figure 6), with detection fractions sharply increasing for stellar masses above $\sim$10$^{10.7}$ M$_{\odot}$. Given the comparatively high [O III] luminosities and stellar masses in our sample, our outflow detection fractions for AGNs of $\sim$ 70 - 80 percent are consistent with other work.  

\subsection{Outflow Properties} \label{subsec: outflow_properties}

\begin{figure}
\includegraphics[width = 0.45\textwidth]{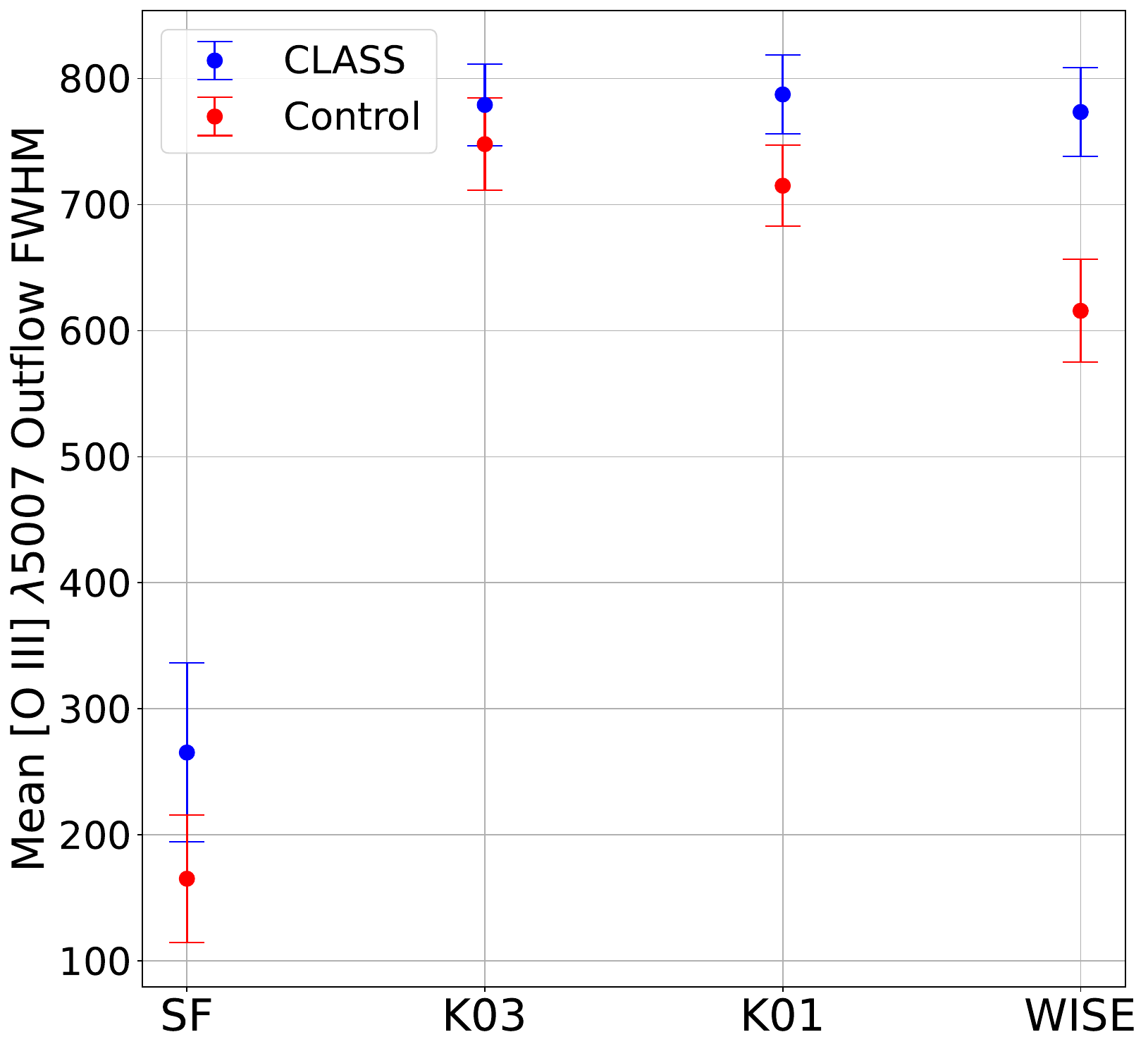}
\caption{FWHM of the [OIII] outflow component in the CL emitters vs. matched control sample for each sub-sample. Error bars are calculated using the standard error of the mean; only the WISE sub-sample exhibits significantly larger outflow FWHM (see Section \ref{subsec: outflow_properties} for details).}
\label{fig:mean_outflow_fwhm}
\end{figure}

Following \citet{Matzko2022}, we use W$_{80}$ of the broad outflowing component to trace the outflow velocity; for Gaussian profiles this is approximately 1.09\texttimes FWHM. For simplicity, we only examine the FWHM directly. In Figure \ref{fig:mean_outflow_fwhm}, we show the average FWHM of the [OIII] outflow component in the CL sample compared to the matched control sample for each sub-sample. Error bars are calculated using the standard error of the mean; we calculate the sample standard deviation and divide it by the square root of the number of galaxies in the sample to obtain the corresponding error bars at the 68\% level. In order to perform a more careful analysis of the distributions of the FWHM of the [OIII] outflow component in CL-emitters and their controls, we also utilize a ks-test and Welch's t-test, implemented by the standard Python library \textsc{scipy.stats}, to determine whether the underlying sample distributions or sample means, respectively, are the same. The latter is used to test the null hypothesis that the means of the two samples (the CLASS and control samples) are identical without assuming equal population variances.\footnote{If we \textit{do} assume equal population variances, our conclusions are unchanged.} The former is used to test the null hypothesis that the underlying cumulative distribution functions (CDFs) of the distributions are identical. We utilize both tests here and throughout the paper. In general, the two tests yield consistent results. As suggested by Figure \ref{fig:mean_outflow_fwhm}, outflows are significantly faster in CL-emitters than their controls in the WISE sub-sample (with corresponding p-values less than 0.0003 for both the ks-test and t-test). The outflow FWHMs of the CL sample for the K01 sub-sample are only marginally insignificant, with p-values $\sim$ 0.054 from the ks-test and t-test.

\begin{figure}
\includegraphics[width = 0.45\textwidth]{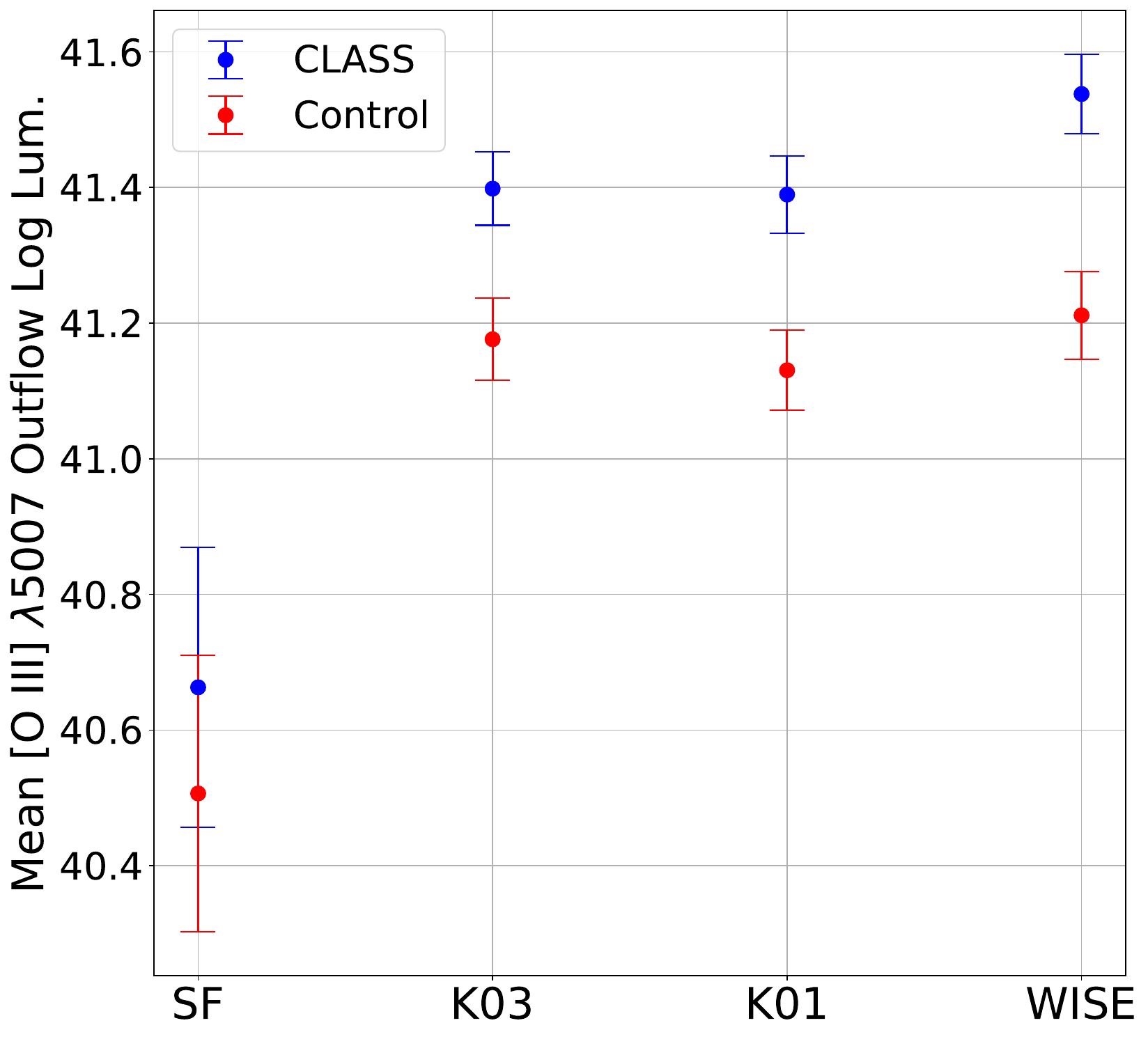}
\caption{[O III] outflow luminosity for each CL-emitter and matched control for each sub-sample. Error bars are calculated using the standard error of the mean; each AGN subsample exhibits significantly higher outflow luminosities than their matched controls (see Section \ref{subsec: outflow_properties} for details).}
\label{fig:mean_outflow_lum}
\end{figure}

In Figure \ref{fig:mean_outflow_lum}, we show the average [OIII] outflow luminosity in the CL sample compared to the matched control sample for each sub-sample. For all of our AGN sub-samples, the outflows are significantly (p-values less than 0.005 for the t-test and ks-test) more luminous in the CL sample compared to their matched controls. We emphasize that this trend is only seen in the luminosity of the outflow component, and not in the luminosity of the narrow [O III] $\lambda$5007 line; by design, the control sample is constructed such that the [O III] $\lambda$5007 luminosity of each CL emitter is matched to the corresponding [O III] $\lambda$5007 luminosity of its control. 

\subsection{Relationship between Outflow Properties and CL Properties} \label{subsec: outflow_cl_relationship}

\begin{figure*}
\gridline{
\fig{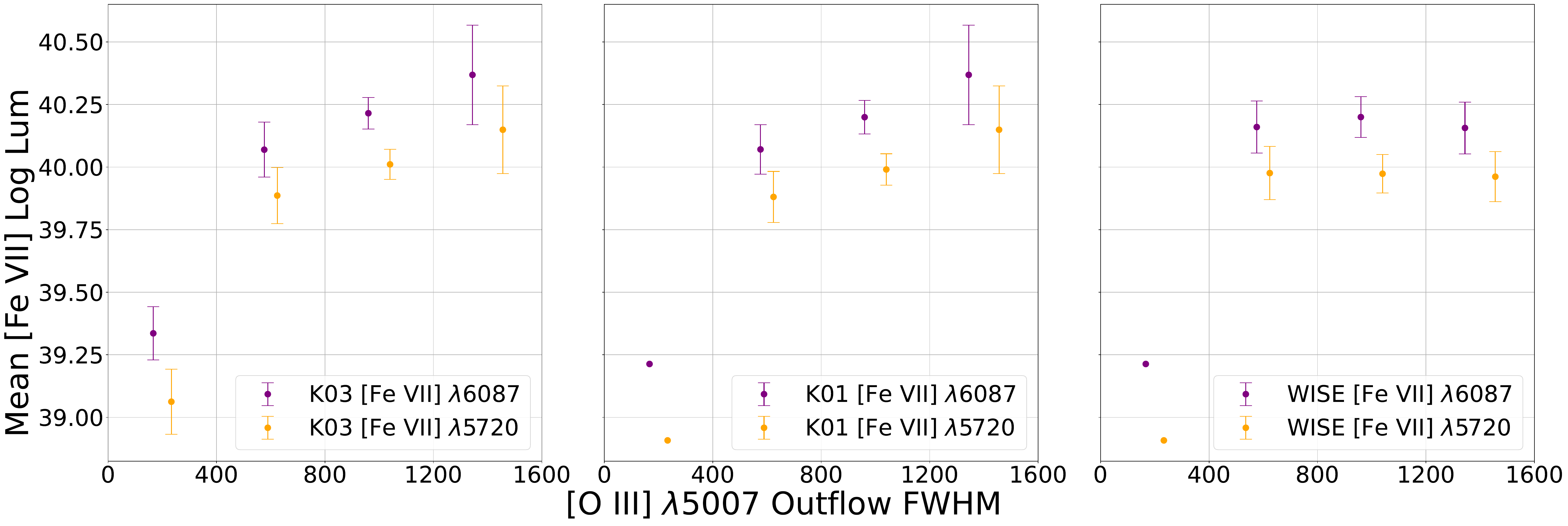}{0.9\textwidth}{}
}
\gridline{
\fig{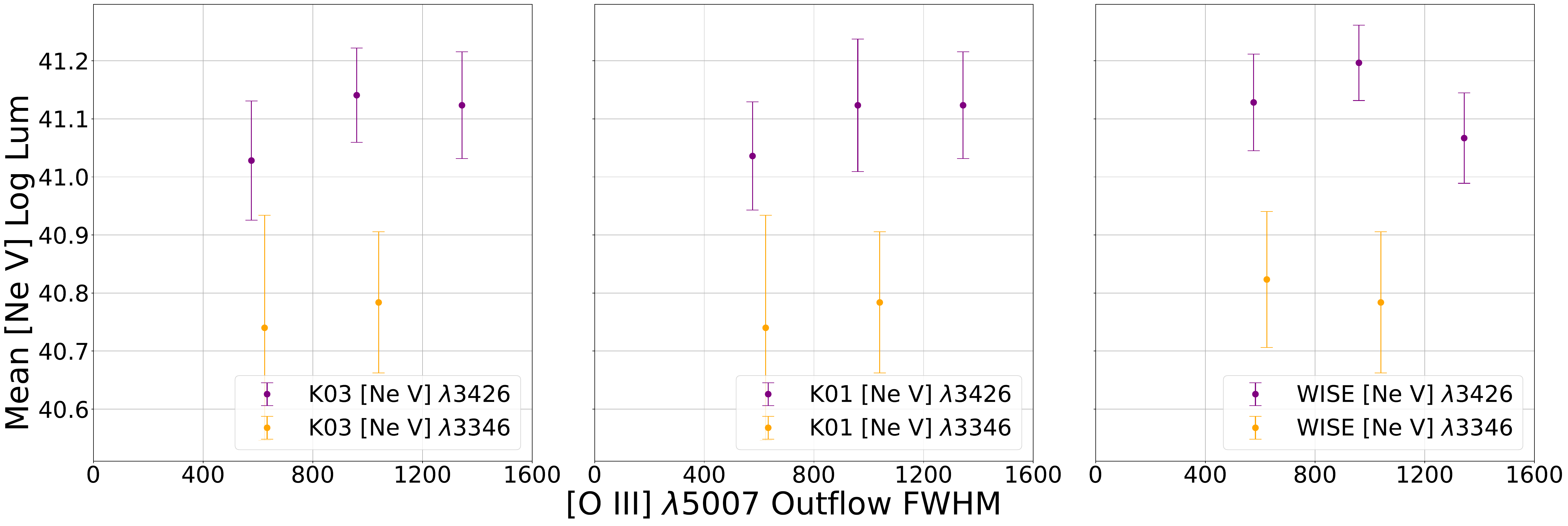}{0.9\textwidth}{}
}
\caption{\textit{Top panel}: [\ion{O}{3}] outflow FWHM vs. the [\ion{Fe}{7}] $\lambda\lambda$6087,5720 average line luminosity for K03, K01, and WISE AGNs, binned by outflow FWHM. \textit{Bottom panel}: Same as the top panel, but with the [\ion{Ne}{5}] $\lambda\lambda$3426,3346 line. Error bars are calculated using the standard error of the mean. Only the K03 and K01 sub-samples for [Fe VII] display significant correlations (see Section \ref{subsec: outflow_properties} for details). }
\label{fig:fwhm_cllum}
\end{figure*}

If ionized outflows enhance CL emission as Figure \ref{fig:incidence} suggests, one might expect a relationship between the ionized outflow properties and the CL properties. To that end, we examine the relationship between outflow properties and CL properties for the [\ion{Ne}{5}] $\lambda\lambda$3426,3346 and [\ion{Fe}{7}] $\lambda\lambda$6087,5720 emission lines, as the limited detections of other CLs in the CLASS sample do not allow for a statistically significant analysis. In the top panel of Figure~\ref{fig:fwhm_cllum}, we plot the relationship between the [\ion{O}{3}] outflow FWHM and the [\ion{Fe}{7}] $\lambda\lambda$6087,5720 line luminosity for K01, K03, and WISE AGNs (the SF sample has an insufficient number of CLs for this analysis). We apply a Spearman rank correlation test to determine if a significant correlation exists between the plotted variables in each panel. We consider a Spearman rank correlation coefficient greater than 0.3 to indicate that a correlation exists between the variables, and a p-value of less than 0.05 to indicate that the correlation is statistically significant. The Spearman rank tests are performed on the underlying data in each figure, not on the displayed mean values. As can be seen in the top panel of Figure \ref{fig:fwhm_cllum}, there is a clear increase in CL luminosity with faster outflows in the K01 and K03 sub-samples for the iron CLs, with a correlation coefficient of $\sim$ 0.4 and p-values less than $\sim$ 0.01. No significant correlation in the WISE sample is seen. In contrast, there is no significant correlation between the CL luminosity with outflow FWHM for the neon lines in any of our AGN sub-samples, as suggested by the bottom panel of figure \ref{fig:fwhm_cllum}. These results are generally expected if CL emission is enhanced in faster outflows through dust destruction in the ionized gas. The increase in iron CL luminosity with outflow velocity might be a consequence of the impact of dust destruction on the gas phase abundance of iron in the ionized gas around AGNs with outflows; iron is depleted onto dust grains in contrast to neon (see Sections \ref{subsec: dust} and \ref{sec: discussion}). 

\par

\begin{figure*}
\gridline{
\fig{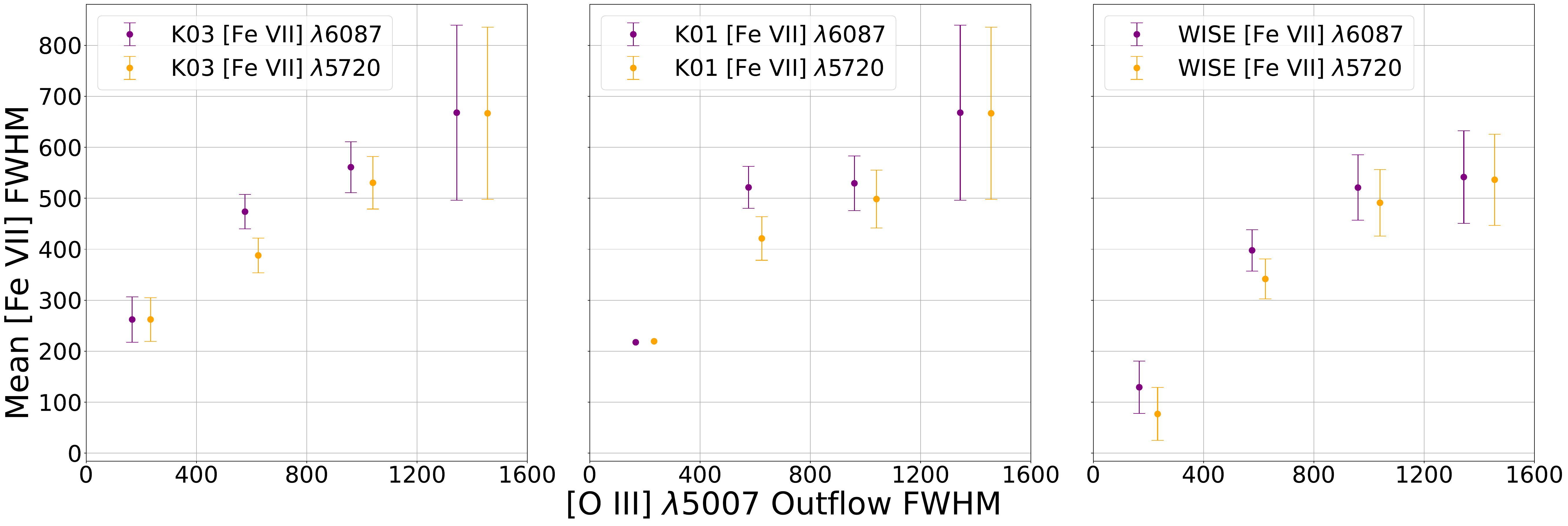}{0.9\textwidth}{}
}
\gridline{
\fig{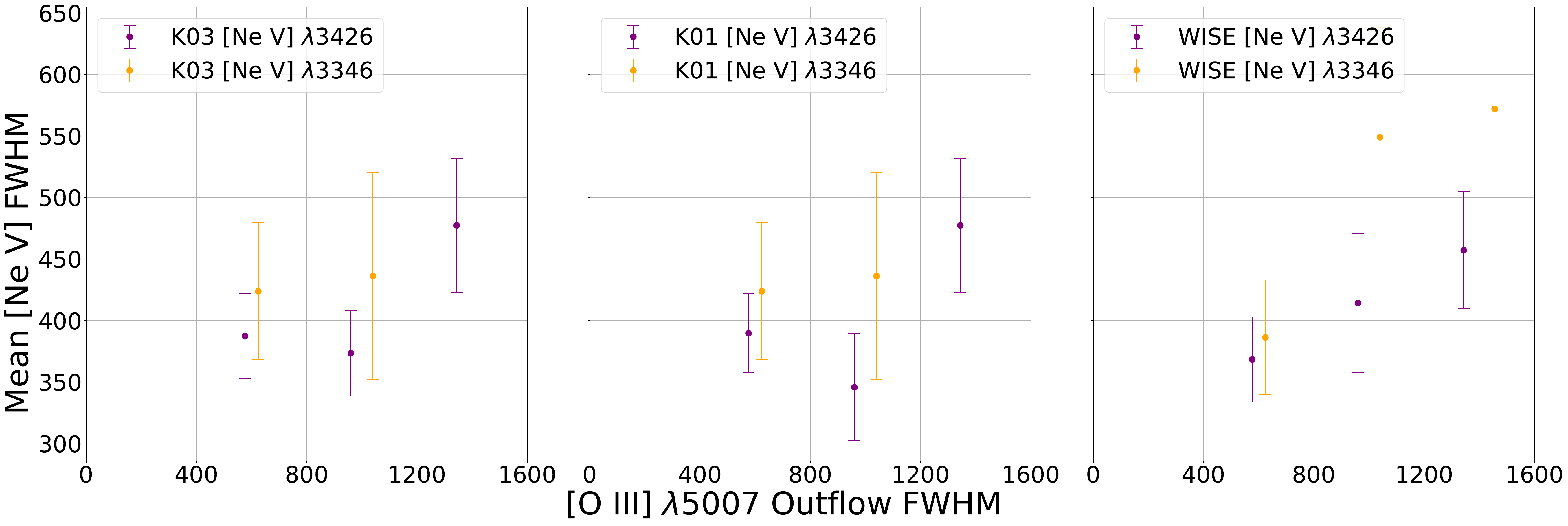}{0.9\textwidth}{}
}
\caption{Average CL FWHM vs. [O III] outflow FWHM for [Fe VII] (top panel) and [Ne V] (bottom panel) CLs, binned in outflow FWHM. Error bars are calculated using the standard error of the mean. A significant relationship is only found for the [Fe VII] lines (see Section \ref{subsec: outflow_properties} for details).}
\label{fig:fwhm_fwhm_trend}
\end{figure*}

In Figure~\ref{fig:fwhm_fwhm_trend}, we plot the relationship between the [\ion{O}{3}] outflow FWHM and the [\ion{Fe}{7}] $\lambda\lambda$6087,5720 FWHM (top panel), as well as the [\ion{Ne}{5}] $\lambda\lambda$3426,3346 FWHM (bottom panel). Here, we generally find that larger outflow FWHMs correspond to larger iron CL FWHMs. This result is statistically significant with a p-value of less than 0.012 and Spearman rank correlation coefficient of $\sim$0.4 for all the iron lines in our AGN sub-samples, except for [Fe VII] $\lambda$6087 in the K01 sample (Spearman rank correlation coefficient $\sim$ 0.2 with p-value of $\sim$0.09). In contrast, such a correlation is not seen even marginally for any of the neon lines. The reasonably clear increase in the FWHM of the Fe CLs with increasing [O III] outflow FWHM suggests that faster outflows traced by the lower ionization [O III] corresponds to CL emission from highly ionized dense gas originating closer to the central engine, as traced by the iron CLs. 

\begin{figure*}
\gridline{
\fig{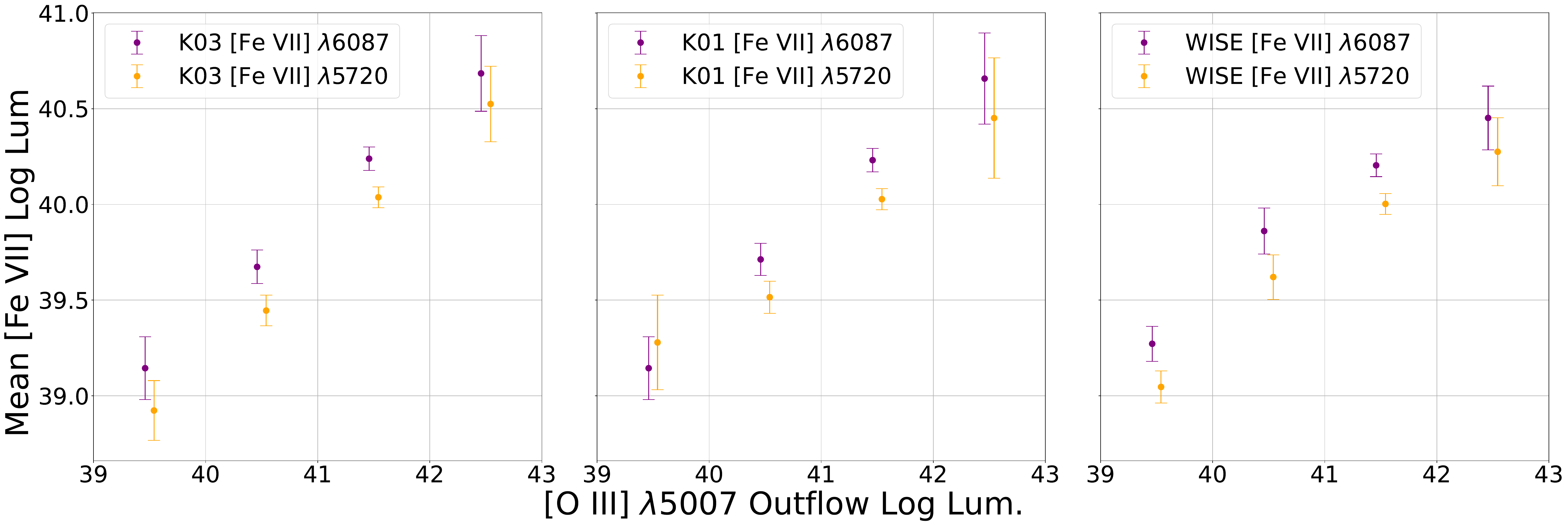}{0.9\textwidth}{}
}
\gridline{
\fig{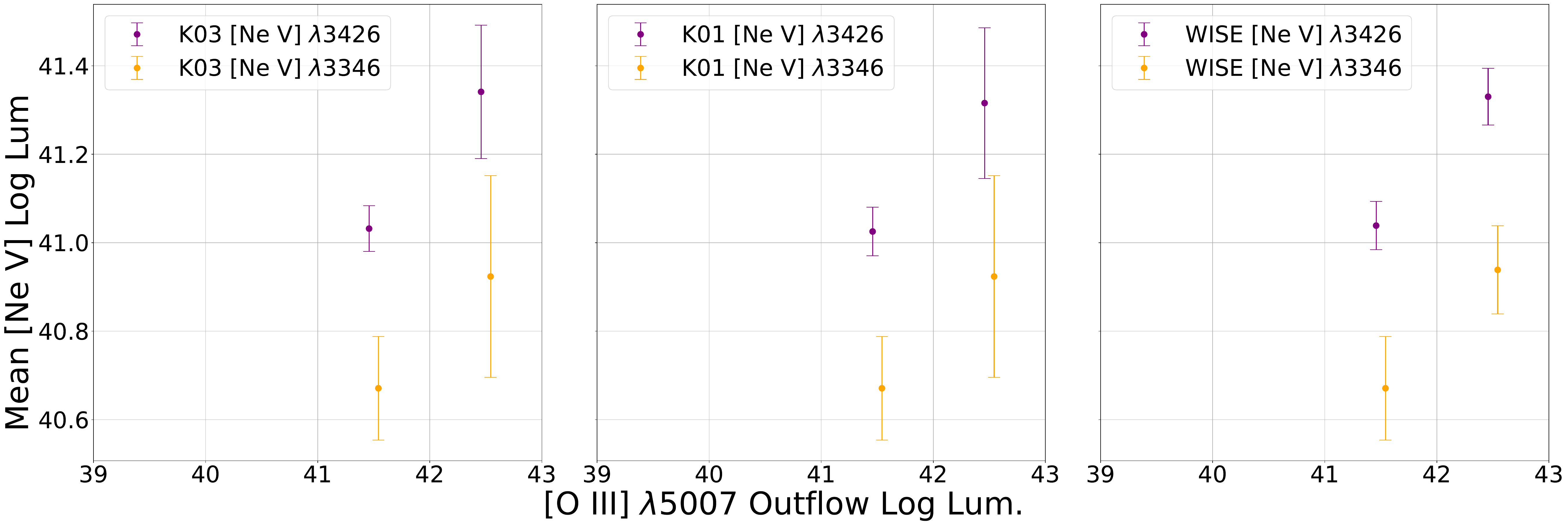}{0.9\textwidth}{}
}
\caption{Average CL luminosity vs. [O III] outflow luminosity for [Fe VII] (top panel) and [Ne V] (bottom panel) CLs. A significant relationship is only found for the [Fe VII] lines (see Section \ref{subsec: outflow_properties} for details).}
\label{fig:lum_lum_trend}
\end{figure*}

Figure \ref{fig:lum_lum_trend} shows the relationship between the [Fe VII] $\lambda \lambda$6087,5720 (top panel) and the [Ne V] $\lambda \lambda$3426,3346 (bottom panel) CL luminosities as a function of the [O III] outflow luminosity. There is a significant increase in the iron line luminosities as the outflow luminosity increases. This holds for both of the plotted iron lines in all three AGN sub-samples with a Spearman correlation coefficient of $\sim$ 0.6 and p-values less than 0.0001. This relationship is, however, generally not significant for the Ne lines (even marginally). Indeed, a significant correlation is only seen in the [Ne V] $\lambda3426$ line in the WISE sample, with a Spearman correlation coefficient of $\sim$ 0.4 and p-value of $\sim$ 0.02. These results echo those found in Figures \ref{fig:fwhm_cllum} and \ref{fig:fwhm_fwhm_trend}, suggesting that stronger outflows are efficient at liberating iron from dust grains and subsequently contributing to an increase in the iron CL emission.  


\subsection{Dust Destruction in the Ionized Gas?} \label{subsec: dust}

\begin{figure}
\includegraphics[width = 0.45\textwidth]{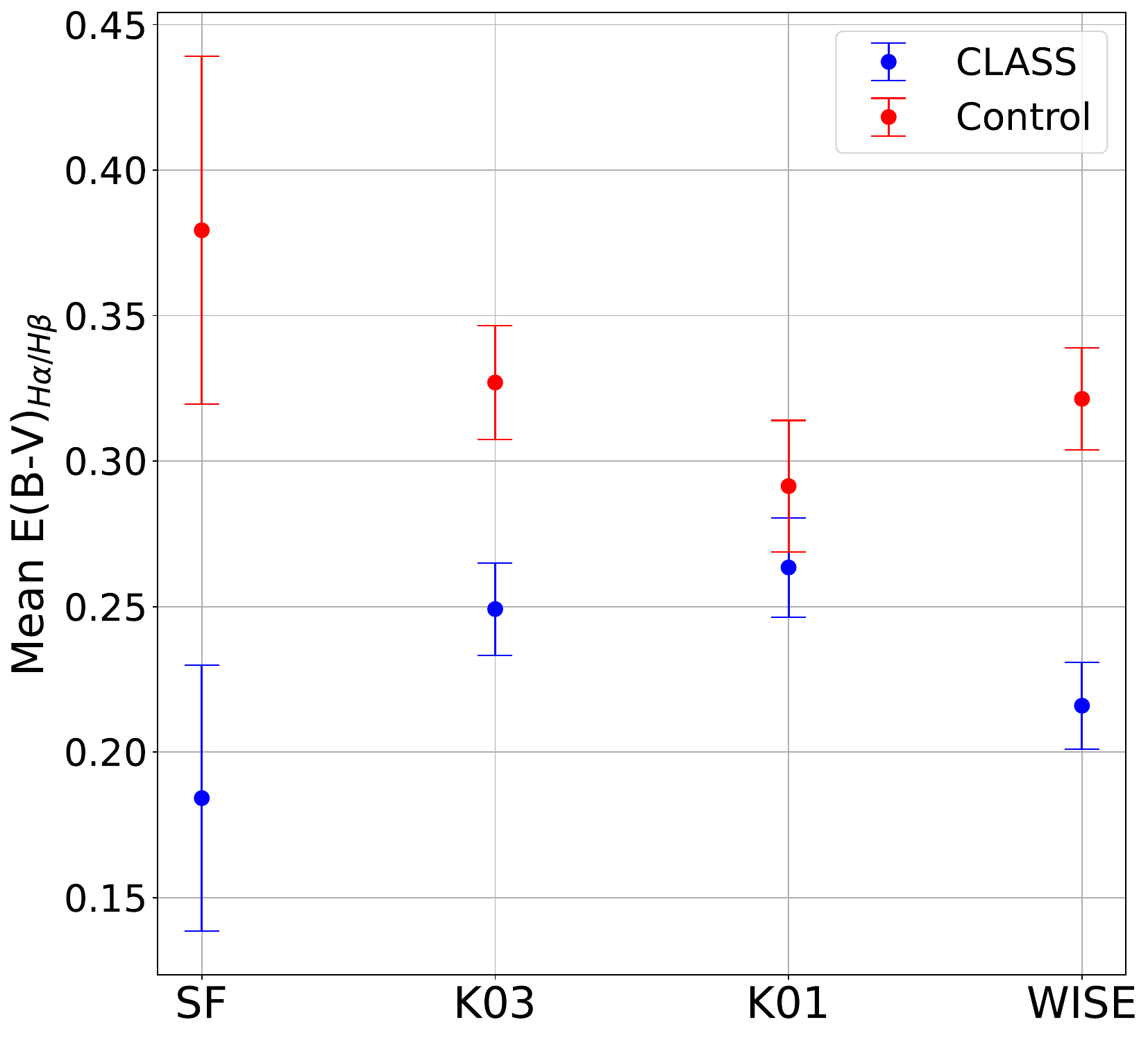}
\caption{Average intrinsic extinction $E(B-V)_{H\alpha/H\beta}$ calculated from the Balmer decrement in CL-emitters vs. matched control sample for each of the sub-samples. Error bars are calculated using the standard error of the mean. The intrinsic extinction is significantly less in the CL-emitters for the SF, K03, and WISE sub-samples, and only marginally significant for the K01 sub-sample (see Section \ref{subsec: outflow_properties} for details).}
\label{fig:mean_ebv}
\end{figure}

\begin{figure*}[]
\includegraphics[width = \textwidth]{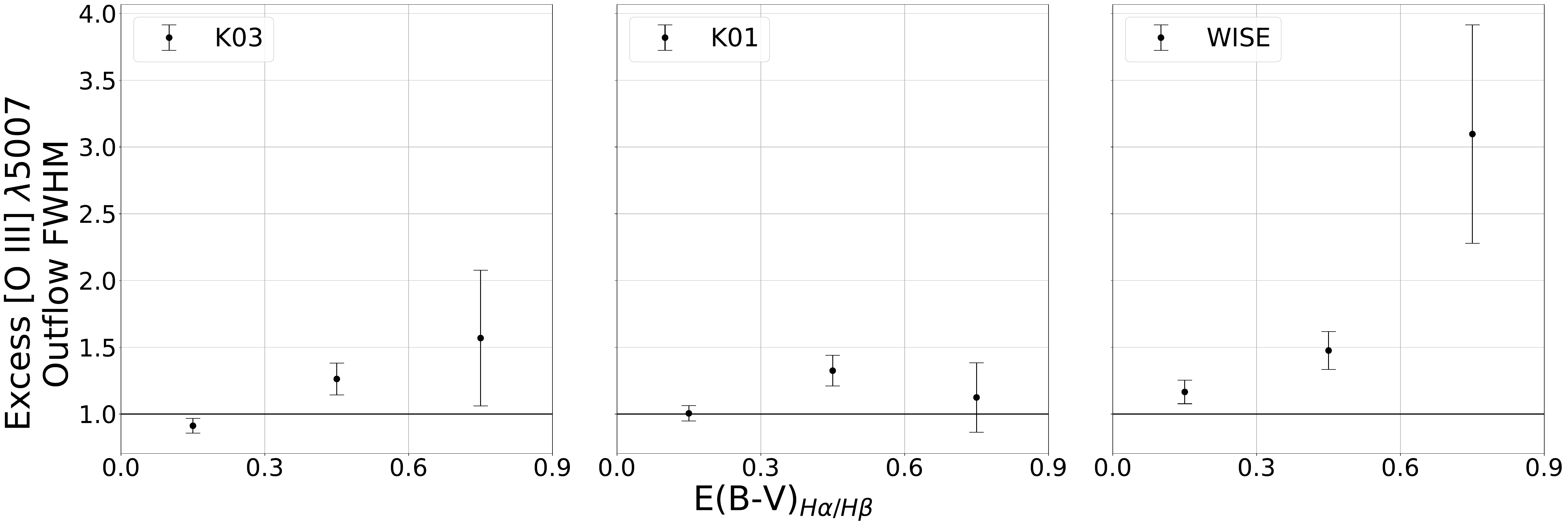}
\caption{`Excess' outflow FWHM; mean outflow FWHM in CL-emitters divided by mean outflow FWHM in controls for each bin of E(B-V). The error bars for the CL-emitters and controls are calculated using the standard error of the mean (see Section \ref{subsec: outflow_properties} for details). }
\label{fig:outflow_fwhm_ebv_excess}
\end{figure*}

If the enhancement of CL emission is related to the presence and properties of outflows due in part to destruction of grains, we might expect CL emitters to have less dust in the ionized gas compared with their controls. In order to investigate the dust content of the galaxies in our sample, we use the Balmer decrement (i.e. the observed flux ratio of H$\alpha$ to H$\beta$) to estimate the intrinsic extinction for each galaxy in our sample. Following \citet{Atek2008}, we calculate the intrinsic extinction via

\begin{equation}
E(B-V)_{H\alpha/H\beta}  = \frac{2.5 \log(R_{\mathrm{int}} / R_{\mathrm{obs}})}{k(\lambda_{\alpha}) - k(\lambda_{\beta})}
\end{equation}
\\
\noindent where $R_{\mathrm{obs}}$ is the observed Balmer decrement, $R_{\mathrm{int}}$ is the intrinsic Balmer decrement assuming Case B recombination, and $k(\lambda_{\alpha}) \sim2.63$ and $k(\lambda_{\beta}) \sim 3.71$ are the extinction curves at the $H\alpha$ and $H\beta$ wavelengths respectively, with values from \citet{Cardelli1989}. For our SF sample, we take $R_{\mathrm{int}} = 2.86$, and for our AGN sample we assume $R_{\mathrm{int}} = 3.1$ due to the harder ionizing radiation field \citep{Osterbrock2006}. 

Figure \ref{fig:mean_ebv} shows the mean intrinsic extinction for each of our sub-samples. As can be seen, the CL galaxies have systematically lower extinction toward the ionized gas, a difference that is most notable in the SF, WISE, and K03 sub-samples. As in Section \ref{subsec: outflow_properties}, we apply a ks-test and Welch's t-test in order to more rigorously assess the statistical differences in the $E(B-V)$ distributions in the CL samples and their matched controls. We find that the $E(B-V)$ distributions for the CL emitters in the SF, K03, and WISE sub-samples are statistically different compared with their controls, with p-values less than 0.003; the K01 AGNs show a less significant difference (p-value of $\sim$0.049 using the ks test; no significant difference seen with the t-test). Thus our statistical analysis demonstrates that CL emitters have lower E(B-V) values than their matched controls. 

\par

If grain destruction is taking place in CL galaxies as a result of outflows, one might expect a relationship between outflow velocity and $E(B-V)$ in CL galaxies relative to their matched controls. In Figure~\ref{fig:outflow_fwhm_ebv_excess} we plot the ''excess'' outflow FWHM, defined as the mean outflow FWHM in the CL galaxies divided by the mean outflow FWHM in the controls as a function of $E(B-V)$. There is a trend of an enhancement in the outflow FWHM with increasing $E(B-V)$ for the K03 and most notably the WISE samples, possibly suggesting that faster outflows are needed for grain  destruction in the CL emitting region from which the CLs arise in galaxies with higher extinction measured through the Balmer decrement. 


\section{Discussion} \label{sec: discussion}
The origin of CLs in dust-free photoionized gas has been noted in a number of previous works. Even as early as 1970, \citet{1970ApJ...161..811N} reported that the gas-phase abundance of Fe in the CL emitting region in local AGNs, estimated using the emission-line flux ratio [Fe VII] $\lambda 6087$ / [Ne V] $\lambda 3426$, is comparable to the solar value, indicating that iron is not depleted in the highly ionized gas from which the emission arises. Photoionization modeling of near-infrared CLs by \citet{1997ApJS..110..287F} also suggested that observed line ratios in AGNs are consistent with photoionized dust free gas, a result also noted by \citet{2003AJ....125.1729N}. In recent work, \citet{McKaig2024} demonstrated that dust significantly impacts coronal line emission, reducing the line luminosity by up to three orders of magnitude compared to dust-free gas. This occurs due to two primary effects: (1) gas-phase depletion onto dust grains suppresses emission line luminosities from the ionized gas, and (2) dust absorbs ionizing radiation, reducing the total line luminosities. Highly refractory elements, such as iron and calcium, are more strongly affected by dust than noble gases like neon, which may explain the relative rarity of CLs from these elements in the observed sample. Given the impact of dust on the CL luminosities, CL emission may be expected to be enhanced in galaxies with reduced dust content, consistent with the finding that CL emitters have lower E(B-V) values relative to their matched controls as discussed in Section \ref{subsec: dust}.

The lower extinction toward the ionized gas in CL emitters relative to the controls is consistent with a scenario in which grain destruction or the removal of grains in the CL emitting region is taking place. The enhancement in the outflow incidence (Figure \ref{fig:incidence}) together with the lower extinction in the CL sample compared to the controls (Figure \ref{fig:mean_ebv}) point to the possibility that outflows may be removing the dust or destroying the grains in the highly ionized gas from which the CL emission arises. Such a scenario is consistent with our finding of higher luminosity outflows in CL emitters relative to their controls (Figure \ref{fig:mean_outflow_lum}) and a correlation between outflow properties with CL properties for the iron CLs, in contrast to the neon lines (Figures \ref{fig:fwhm_cllum},\ref{fig:fwhm_fwhm_trend},\ref{fig:lum_lum_trend}).


With the advent of high sensitivity and high spatial resolution integral field observations, a more detailed exploration of the relationship between outflows and the presence and distribution of CL emission, the dust extinction distribution, and the gas phase metallicity can be conducted. In recent work, spatially resolved IFU observations with JWST/NIRSpec of outflows traced by the [O III] $\lambda$5007 emission in high-z galaxies have shown that peaks in the outflow velocity are spatially-coincident with increases in the gas-phase metallicity and lower dust extinction compared to other regions in the galaxies \citep[][]{Parlanti2024, Pino2024}, strongly suggesting that ionized outflows are indeed capable of liberating the highly refractory elements from dust grains into the gas phase. 

We note that the interplay between dust, outflows, and CL emission is likely complex. Various modeling efforts have suggested that dust is a crucial ingredient needed to provide sufficient radiation pressure to drive strong outflows \citep[e.g.][]{Ishibahi2015,Ishibashi2016,Costa2018,Arakawa2022}. Indeed, the most powerful outflows have been found in dust obscured quasars , where [O III] outflow velocities are seen in excess of 6000 km~s$^{-1}$ \citep{Perrotta2019}, highlighting the important role of dust in AGN feedback processes. While [O III] outflows trace more extended (and possibly dusty) gas, the CLs may originate in distinctly different, dust-free, regions that are closer to the central AGN \citep{Mullaney2008}. Several works consider the possibility that efficient CL emission arises from an X-ray-heated outflowing wind from the inner wall of a dusty torus, where photoevaporation of dust grains is taking place \citep{1995ApJ...450..628P, Dorodnitsyn2008,Mullaney2008}. Indeed, \citet{Glidden2016} model the strength of iron CL emission for various viewing angles into an obscuring structure around an AGN, and find their models broadly agree with observed CL fluxes from Coronal Line Forest AGN \citep{Rose2015}. In this scenario, CL-emitting gas can be created and launched from the inner wall of a dusty torus in a radiatively driven wind while iron-carrying dust grains are sublimated \citep{Mullaney2009}. If additional winds are present, the CL-emitting clouds may be carried to greater spatial extents where the bulk of the [O III] emission rises.

Interestingly, as noted in Section \ref{sec: intro}, some LRDs appear to be missing a hot dust component expected for an AGN torus \citep{2024ApJ...968....4P,2024ApJ...968...34W}, which may contribute to the lack of CL emission seen in some high-z JWST targets \citep{2023ApJ...954L...4K,2023A&A...677A.145U,2024arXiv240913047L}.

It is well known that AGN feedback is an important ingredient in galaxy evolution \citep[for detailed reviews see][]{Fabian2012, Morganti2017, Harrison2024}. While many simulations highlight the necessity of AGN feedback in galaxy evolution \citep[e.g.][]{Glines2020, Tillman2023}, they can still struggle to replicate finer observational details \citep[][]{Monaco2007, Fontanot2007, Contini2024}, with certain modes of AGN feedback found to be unable to match observational constraints \citep[][]{Su2020, Glines2020, Su2021}. Nonetheless, \citet{Torrey2020} model the coupling of fast nuclear winds to the host galaxy's ISM in isolated disc galaxies. These models showed that low, and even moderate, luminosity winds from AGNs are inefficient at coupling to the ambient ISM and are unlikely to have a significant large-scale impact on the host galaxy. However, sufficiently luminous AGN may have nuclear winds that are capable of having a significant impact on the host galaxy's ISM, particularly in the case of thicker gas disks \citep{Sivasankaran2025}. Although speculative, our observation of an increased ionized outflow incidence in the presence of CL emission could be tracing a particular `phase' of AGN feedback, where we are seeing more efficient coupling between nuclear winds and larger-scale outflows that are able to have a more significant impact on the ISM of the host galaxy.

\section{Conclusion} \label{sec: conclusion}


We have presented the first systematic study of the incidence and properties of ionized outflows traced by the [O III] $\lambda$5007 emission line in a sample of CL-emitting galaxies relative to a carefully constructed control sample of galaxies without CLs. Our main results are as follows:

\begin{enumerate}
   \item CL-emitters that are identified as AGN through optical or infrared diagnostics systematically display an increased prevalence of [O III] outflows compared to their matched controls. The difference in outflow incidence in the CL-emitters and controls ($\sim$ 80\% vs 65\%) is significant to the 90\% level (Figure \ref{fig:incidence}). 
   \item For each AGN sub-sample, outflows are significantly more luminous in galaxies that display CLs. Since the control samples are matched in [O III] luminosity, this is not simply a result of CL-emitters being intrinsically brighter than the controls (Figure \ref{fig:mean_outflow_lum}). 
   \item We find significant correlations between the luminosity of the iron CLs and [O III] outflow velocity. We emphasize this trend is only seen in the iron CLs, and not in the neon CLs. Similar relationships are found between the CL FWHM and [O III] outflow luminosity (Figures \ref{fig:fwhm_cllum}, \ref{fig:fwhm_fwhm_trend}, \ref{fig:lum_lum_trend}).
   \item The intrinsic extinction, traced by the Balmer decrement, is systematically lower in the CL-emitters compared to their matched controls (Figure \ref{fig:mean_ebv}). This is highly suggestive of an enhancement of CL emission with dust destruction in an outflowing wind. 
    
\end{enumerate}

Our results are consistent with the scenario proposed by \citet{1995ApJ...450..628P} in which dust destruction in an outflowing wind from the inner edge of a torus structure is taking place, causing efficient CL emission. Subsequent winds or outflows may propel the highly ionized grain-free gas to larger spatial scales. Hence, these CL-emitters may be capturing a distinct phase in AGN feedback. Follow-up observations with spatially resolved IFU data in CL-emitters with outflows (ideally with multi-phase outflows) would provide valuable insight into the dynamics of outflowing gas at various spatial scales and its  interaction with the host galaxy ISM. 

\section{Acknowledgements}

The computational work carried out in this work were run on ARGO and HOPPER, research computing clusters provided by the Office of Research Computing at George Mason University, VA. (\url{ http://orc.gmu.edu})

This research made use of Astropy,\footnote{\url{http://www.astropy.org}} a community-developed core Python package for Astronomy \citep{2013A&A...558A..33A}.  

 Funding for SDSS-III has been provided by the Alfred P. Sloan Foundation, the Participating Institutions, the National Science Foundation, and the U.S. Department of Energy Office of Science. The SDSS-III web site is \href{http://www.sdss3.org/}{http://www.sdss3.org/}.

 SDSS-III is managed by the Astrophysical Research Consortium for the Participating Institutions of the SDSS-III Collaboration including the University of Arizona, the Brazilian Participation Group, Brookhaven National Laboratory, Carnegie Mellon University, University of Florida, the French Participation Group, the German Participation Group, Harvard University, the Instituto de Astrofisica de Canarias, the Michigan State/Notre Dame/JINA Participation Group, Johns Hopkins University, Lawrence Berkeley National Laboratory, Max Planck Institute for Astrophysics, Max Planck Institute for Extraterrestrial Physics, New Mexico State University, New York University, Ohio State University, Pennsylvania State University, University of Portsmouth, Princeton University, the Spanish Participation Group, University of Tokyo, University of Utah, Vanderbilt University, University of Virginia, University of Washington, and Yale University. 

This publication makes use of data products from the Wide-field Infrared Survey Explorer, which is a joint project of the University of California, Los Angeles, and the Jet Propulsion Laboratory/California Institute of Technology, and NEOWISE, which is a project of the Jet Propulsion Laboratory/California Institute of Technology. WISE and NEOWISE are funded by the National Aeronautics and Space Administration.

\bibliography{bib}{}
\bibliographystyle{aasjournal}

\end{document}